\documentclass[preprint,amsmath,amssymb,prb,floatfix,superscriptaddress]{revtex4}

\usepackage[dvips]{graphicx}
\usepackage{dcolumn}
\usepackage{bm}
\usepackage{mathrsfs}

\begin{document}
\title{Measurement of quadratic Terahertz optical nonlinearities using second harmonic lock-in detection}
\author{Shuai Lin}
 \affiliation{Department of Physics and Engineering Physics, Tulane University, 6400 Freret St., New Orleans, LA 70118, USA}
\author{Shukai Yu}
 \affiliation{Department of Physics and Engineering Physics, Tulane University, 6400 Freret St., New Orleans, LA 70118, USA}
\author{Diyar Talbayev}
 \email{dtalbayev@gmail.com}
 \affiliation{Department of Physics and Engineering Physics, Tulane University, 6400 Freret St., New Orleans, LA 70118, USA}

\date{\today}

\newcommand{\fzmo}{FeZnMo$_3$O$_8$}
\newcommand{\cm}{\:\mathrm{cm}^{-1}}
\newcommand{\T}{\:\mathrm{T}}
\newcommand{\mc}{\:\mu\mathrm{m}}
\newcommand{\ve}{\varepsilon}
\newcommand{\dg}{^\mathtt{o}}

\begin{abstract}
We present a method to measure quadratic Terahertz optical nonlinearities in Terahertz time-domain spectroscopy.  We use a rotating linear polarizer (a polarizing chopper) to modulate the amplitude of the incident THz pulse train.  We use a phase-sensitive lock-in detection at the fundamental and the second harmonic of the modulation frequency to separate the materials' responses that are linear and quadratic in Terahertz electric field.  We demonstrate this method by measuring the quadratic Terahertz Kerr effect in the presence of the much stronger linear electro-optic effect in the (110) GaP crystal.  We propose that the method can be used to detect Terahertz second harmonic generation in noncentrosymmetric media in time-domain spectroscopy, with broad potential applications in nonlinear Terahertz photonics and related technology.
\end{abstract}

\maketitle

\section{Introduction}
Nonlinear Terahertz (THz) optics has blossomed into an exciting and active area of research with the advent of table-top high-field THz sources\cite{hebling:b6, hoffmann:a29, hirori:091106, hwang:1447, kampfrath:680}.  Among a multitude of nonlinear phenomena, we focus here on the nonlinearities that are quadratic in THz electric field $E^T$.  Such quadratic effects can be induced in the second and third orders via the nonlinear polarizabilities $\chi^{(2)}(E^T)^2$ and $\chi^{(3)}(E^T)^2E^{\omega}$, where $E^{\omega}$ is an optical field.  The second order nonlinearity $\chi^{(2)}$ can lead to second harmonic generation (or sum frequency mixing) in noncentrosymmetric crystals.  While it appears that the peak electric field strength available from the table-top THz sources may be sufficient to generate the second harmonic of the fundamental THz pulse\cite{merbold:7262}, THz second harmonic generation has not yet been reported.  For a typical time-domain single-cycle fundamental THz pulse, the generated second harmonic THz pulse will have a significant time-domain and spectral overlap with the fundamental THz pulse.  This makes it difficult to distinguish the fundamental and second harmonic THz waves in a measurement, as the second harmonic amplitude may be several orders of magnitude smaller than the fundamental\cite{merbold:7262}. 

The third order nonlinearity $\chi^{(3)}$ results in a THz-induced optical birefringence proportional to $(E^T)^2$, also termed THz Kerr effect.  The birefringence can be detected as polarization rotation of the optical gating beam $E^{\omega}$ in a measurement.  The first observation of the THz Kerr effect was reported by Hoffmann $et$ $al.$ in several liquids\cite{hoffmann:231105}.  Following their observation, THz Kerr effect was investigated in many other liquid and solid materials\cite{zalkovskij:221102, cornet:1648, allodi:234204, sajadi:28985, sarbak:26749, shalaby:036106, bodrov:084507, sajadi:14963, kampfrath:319, kampfrath:1279}.  Up to now, the reported observations of the THz Kerr effect occurred in liquids or crystals in geometries where the quadratic THz-induced birefringence is unobscured by the linear electro-optic (Pockels) effect.  The latter effect results from the second-order nonlinearity and provides a common means for time-domain detection of THz electric field using femtosecond optical pulses - the electro-optic sampling\cite{wu:2924,nahata:2321,wu:1784}.  When both electro-optic effects are present, the linear birefringence due to Pockels effect is usually much stronger that the quadratic birefringence due to THz Kerr effect.  Measuring and characterizing the much weaker THz Kerr effect in the presence of the stronger Pockels effect is a difficult experimental question.

\begin{figure}[ht]
\begin{center}
\includegraphics[width=3.5in]{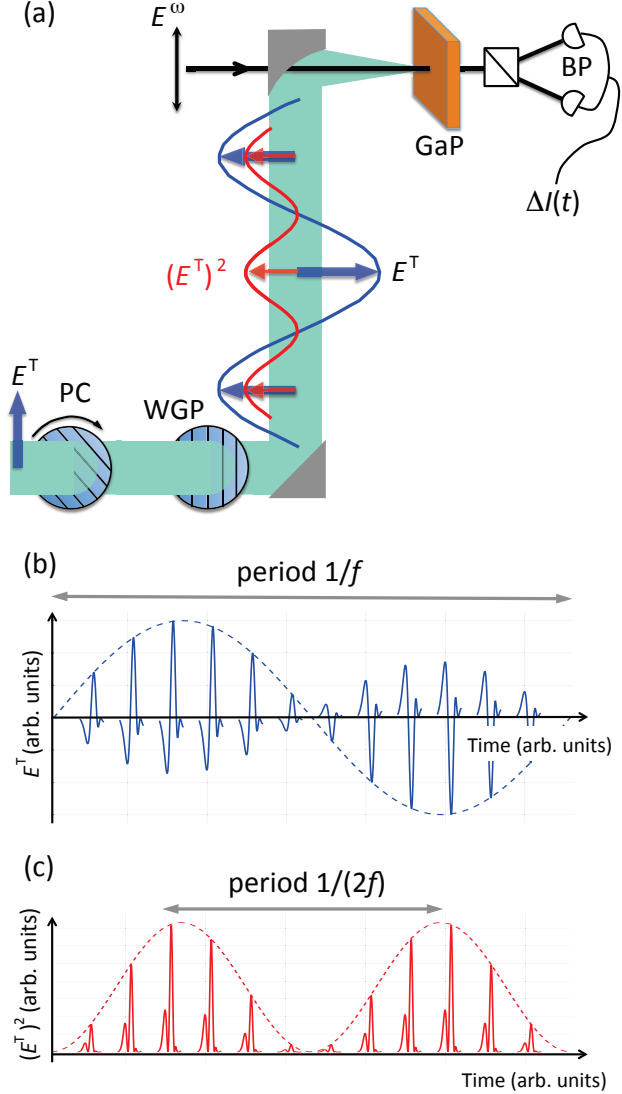}
\caption{\label{fig:fig1} (a) Schematic drawing of the second harmonic lock-in detection experiment.  The LiNbO$_3$ source of THz pulses $E^T$ is not shown.  A 1-kHz train of vertically polarized THz pulses passes through the polarizing chopper (PC) and a fixed wire-grid polarizer (WGP) with a horizontal transmission axis.  After the WGP, the THz pulse train is horizontally polarized and sinusoidally modulated at frequency $f$ (the blue sinusoid).  The product $(E^T)^2$ is modulated at the frequency 2$f$ (the red sinusoid).  The THz pulse train impinges on the (110)-oriented GaP receiver for detection by the vertically polarized gating beam $E^{\omega}$.  Balanced photodiodes (BP) measure the THz-induced birefringence in GaP.  (b) Illustration of the sinusoidally modulated THz pulse train after the WGP.  The polarization is horizontal and the strength of the THz electric field changes sinusoidally over one period.  The THz pulses change polarity from positive to negative.  The duration of THz pulses and the pulse separation are not drawn to scale.  (c) Sinusoidal modulation of the product $(E^T)^2$.  The modulation frequency is 2$f$. }
\end{center}
\end{figure}

In this article, we present a method to measure quadratic THz optical nonlinearities that scale as $(E^T)^2$.  We illustrate the method by measuring the THz Kerr effect in a (110)-oriented GaP crystal via the optical birefringence due to the third order nonlinear polarizability $\chi^{(3)}(E^T)^2E^{\omega}$, where $E^{\omega}$ is the electric field of the optical gating beam.  Importantly, the (110)-oriented GaP crystal also exhibits a much stronger birefringence due to the linear electro-optic Pockels effect governed by the second order polarizability $\chi^{(2)}E^TE^{\omega}$.   We show that our method allows the separation and reliable measurement of the two effects, one of which is quadratic and the other is linear in THz field $E^T$.  We also show how this method can be extended to the time-domain detection of the THz second harmonic generation due to the second order polarization of the form $\chi^{(2)}(E^T)^2$.  Our method shows a strong potential for broad adoption and applications in nonlinear THz photonics.  

\section{Experimental results and discussion}
\begin{figure}[ht]
\begin{center}
\includegraphics[width=3.5in]{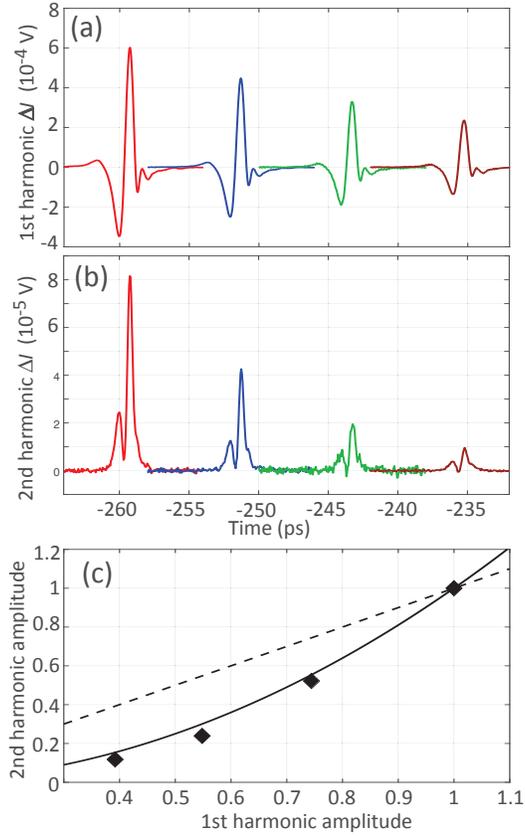}
\caption{\label{fig:tdsgap} (a) Time-domain THz pulses $E^T$ measured using linear electro-optic sampling and the first harmonic detection.  The pulses of different amplitude are obtained by inserting high-resistivity Si wafers in the THz bean path.  The number of wafers varies from 0 to 3.  (b) Time-domain $(E^T)^2$ spectra measured using the second harmonic lock-in detection and quadratic electro-optic effect.  The colors of the spectra correspond to the same color spectra in panel (a).  (c) Relative peak amplitudes of the first and second harmonic spectra from panels (a) and (b).  Diamond symbols - measured amplitudes.  Dashed line - linear function.  Solid line - quadratic function.}
\end{center}
\end{figure}

Our measurement uses a THz time-domain spectrometer based on a 1-kHz repetition rate regenerative amplifier.  We use optical rectification and tilted-wavefront phase matching in a LiNbO$_3$ prism to generate high-field THz pulses.  We use linear electro-optic sampling in a (110)-oriented GaP crystal for THz detection\cite{wu:2924,nahata:2321,wu:1784} and estimate that the peak electric field incident on GaP is over 100 kV/cm \cite{yu:125201}.   The 1-kHz train of vertically polarized THz pulses from LiNbO$_3$ is modulated by a continuously rotating wire-grid polarizer - the polarizing chopper\cite{aschaffenburg:241114,morris:12303}.  The THz pulse train then passes through a fixed wire-grid polarizer with a horizontal transmission axis, Fig.~\ref{fig:fig1}(a).  This arrangement results in a sinusoidal modulation of the THz pulse train in which the polarity of the THz pulse switches from positive to negative, Fig.~\ref{fig:fig1}(b).   The sinusoidal modulation frequency is $f$=77 Hz.  The THz field incident on the (110) GaP receiver is horizontally polarized and the gating optical beam is vertically polarized.  The time-domain amplitude of the THz pulse is measured using balanced photodiodes and a lock-in amplifier with $f$ as the reference frequency, Fig.~\ref{fig:fig1}(a).  The measured quantity $\Delta I(t)$ is the difference in the gating beam intensity on the balanced photodiodes as a function of the time $t$ between the gating optical pulse and the THz field.  $\Delta I(t)$ is proportional to the linear THz-induced birefringence due to the electro-optic Pockels effect\cite{wu:2924,nahata:2321,wu:1784}.  This method is commonly known as electro-optic sampling.  Figure~\ref{fig:tdsgap}(a) shows the measured THz field pulses in time domain. 

The same setup also allows the measurement of the THz Kerr effect that scales as $(E^T)^2$.  The product $(E^T)^2$ is modulated by the polarizing chopper at twice the fundamental frequency, 2$f$, as illustrated in Fig.~\ref{fig:fig1}(c).  By detecting the quantity $\Delta I(t)$ at the second harmonic of the fundamental modulation frequency (2$f$=154 Hz), we measure the THz Kerr effect in (110) GaP.  Figure~\ref{fig:tdsgap}(b) displays the time-domain $(E^T)^2$ spectra measured this way.  The colors of THz Kerr spectra in Fig.~\ref{fig:tdsgap}(b) correspond to the same-color linear THz spectra in Fig.~\ref{fig:tdsgap}(a).  We emphasize that the measurements in Figs.~\ref{fig:tdsgap}(a,b) were taken under the same experimental conditions and geometry.  The only difference between the measurements is the lock-in amplifier detection at the fundamental and the second harmonic frequency of the polarizing chopper modulation.  The THz amplitude in Fig.~\ref{fig:tdsgap}(a) was varied by inserting high-resistivity Si wafers in the THz beam path; the number of wafers ranged from 0 to 3.  Figure~\ref{fig:tdsgap}(c) quantifies the peak amplitude of $(E^T)^2$ as a function of the peak amplitude of $E^T$; the highest peak amplitude of each quantity is normalized to 1 in the Figure.  As expected, we observe that the relationship is quadratic.  We note that the second harmonic lock-in detection allows to directly measure the quadratic THz Kerr birefringence in the presence of linear electro-optic Pockels birefringence that is more than 20 times stronger.  

\begin{figure}[ht]
\begin{center}
\includegraphics[width=3.5in]{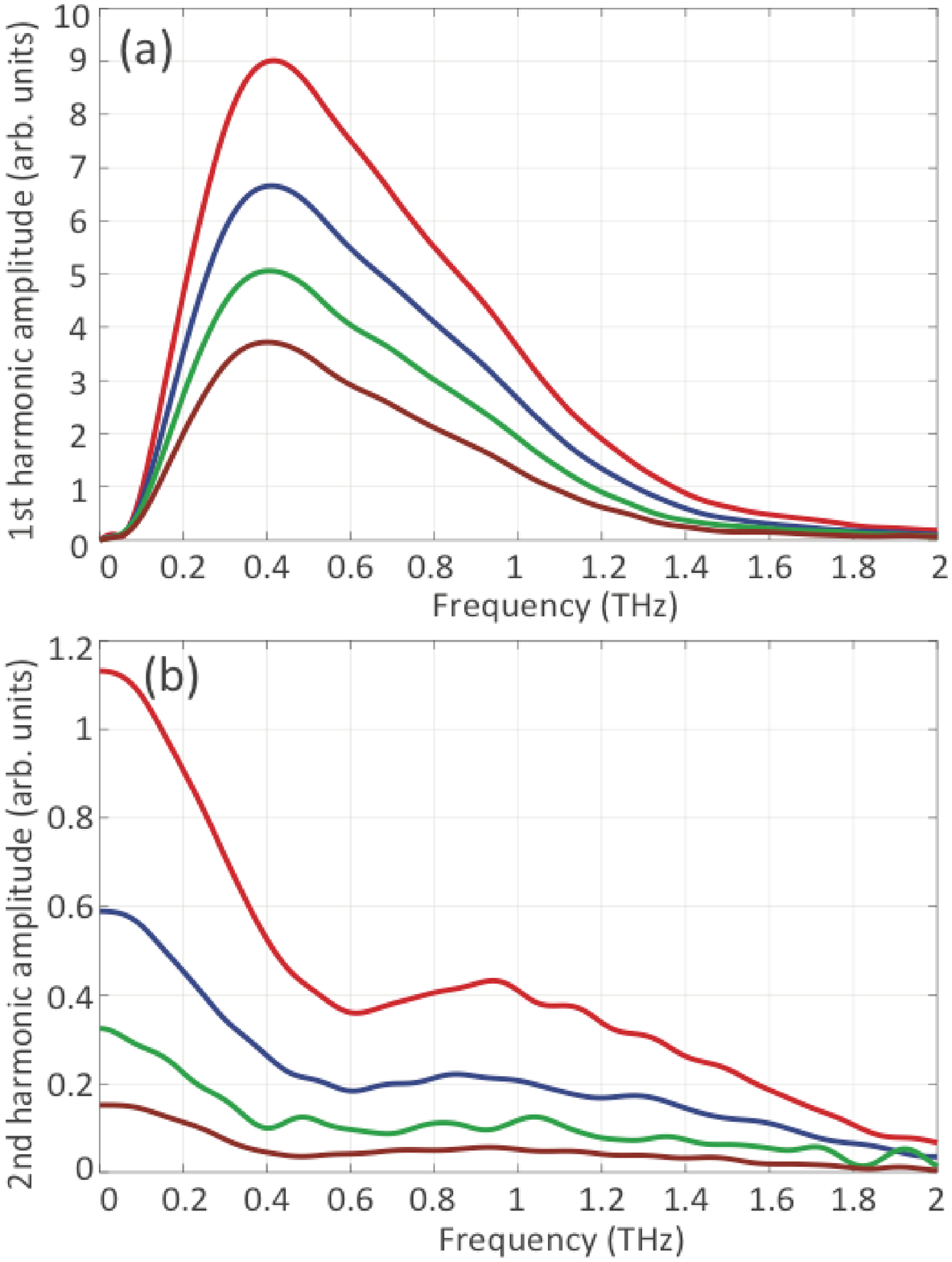}
\caption{\label{fig:fdsgap} (a) Frequency-domain amplitudes for the first-harmonic time-domain spectra from Fig.~\ref{fig:tdsgap}(a). (b) Frequency-domain amplitudes for the second-harmonic time-domain spectra from Fig.~\ref{fig:tdsgap}(b).}
\end{center}
\end{figure}

Figure~\ref{fig:fdsgap} compares the frequency content of the linear and quadratic THz spectra measured via the first and second harmonic lock-in detection.  The frequency-domain spectra in the figure display the Fourier transform amplitude of the spectra in Figs.~\ref{fig:tdsgap}(a,b).  Figure~\ref{fig:fdsgap}(a) shows typical linear frequency-domain THz spectra with near identical frequency content for the different THz peak amplitudes incident on the GaP receiver.  The linear frequency-domain spectra peak just above 0.4 THz.  The quadratic, second harmonic THz spectra peak around 0.8-0.9 THz, twice the frequency of the linear spectrum frequency peak.  The second harmonic spectra also possess a large zero-frequency component because $(E^T)^2$ in time domain has no negative components.  For comparison, we computed $(E^T)^2$ and its Fourier transform from the measured linear $E^T$.  In these numerically computed spectra, the zero-frequency component always has twice the amplitude of the high-frequency peak.  The experimentally measured $(E^T)^2$ spectra only roughly follow this trend, Fig.~\ref{fig:fdsgap}(b). 

\begin{figure}[ht]
\begin{center}
\includegraphics[width=3.5in]{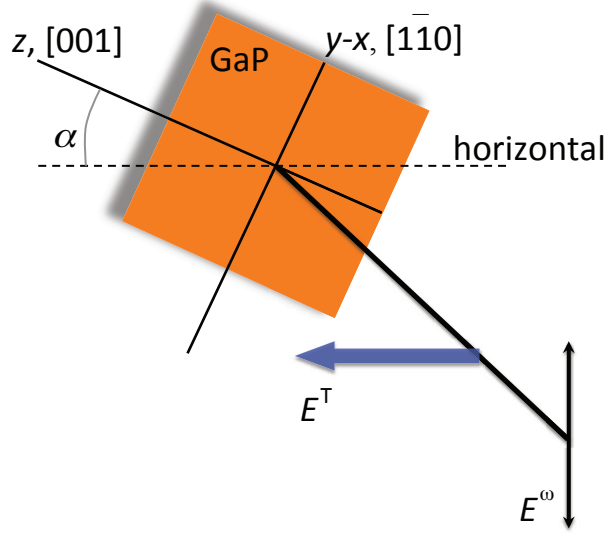}
\caption{\label{fig:fig4} Geometry of the linear electro-optic and THz Kerr birefringence measurements on the (110) GaP crystal.  The peak THz electric field $E^T$ is horizontal.  The optical gating beam $E^\omega$ is polarized vertically.  $\alpha$ labels the angle between the [001] $z$ axis of the crystal and the horizontal.  The other crystalline axes are [100] $x$ and [010] $y$.}
\end{center}
\end{figure}

The data presented in Figs.~\ref{fig:tdsgap} and~\ref{fig:fdsgap} provide strong evidence that the second harmonic lock-in detection measures the quadratic THz nonlinearity due to the THz Kerr effect.  The induced THz Kerr birefringence is described by the third order nonlinear polarization
\begin{equation}
\label{eq:p3}
P^{(3)}_i=\chi^{(3)}_{ijkl}E^T_jE^T_kE^{\omega}_l,
\end{equation}
where the cartesian coordinates $i,j,k,l=1$, $2$ or $3$ and a summation over repeated indices is implied.  The third-order nonlinear tensor $\chi^{(3)}_{ijkl}$ has 21 non-zero elements of which 4 are independent\cite{boyd:nonlinearoptics}:
\begin{eqnarray}
\label{eq:chi3}
\begin{split}
xxxx=yyyy=zzzz,\\
yyzz=zzyy=zzxx=xxzz=xxyy=yyxx,\\
yzyz=zyzy=zxzx=xzxz=xyxy=yxyx,\\
yzzy=zyyz=zxxz=xzzx=xyyx=yxxy.
\end{split}
\end{eqnarray}
Equations (\ref{eq:p3}) and (\ref{eq:chi3}) allow the modeling of THz Kerr birefringence as a function of the angular orientation of the (110) GaP crystal.  For a complete model of such angular dependence, the knowledge of the four independent nonlinear parameters in Eq. (\ref{eq:chi3}) is necessary.  In the absence of such knowledge, we can compute the THz Kerr birefringence and the balanced photodiode current $\Delta I(t)$ for the two special cases $\alpha$=$0\dg$ and $\alpha$=$90\dg$.  We define $\alpha$ as the angle between the peak THz electric field vector and the in-plane [001] direction of the GaP crystal.  We use the cartesian coordinate system associated with the GaP crystal and shown in Fig.~\ref{fig:fig4}.  The axes $x$, $y$, and $z$ correspond to the cubic [100], [010], and [001] directions in this system.

In the case $\alpha$=$0\dg$, we set
\begin{eqnarray}
\nonumber
\vec{E}^T=E^T_0\left( \begin{matrix}0\\ 0\\ 1\\ \end{matrix} \right) \; \textrm{and} \quad \vec{E}^{\omega}=\frac{E^{\omega}_0}{\sqrt{2}}\left( \begin{matrix}-1\\ 1\\ 0\\ \end{matrix}\right)\\ 
\nonumber
\textrm{and obtain} \quad
\vec{P}^{(3)}=\left(
                              \begin{matrix}\chi^{(3)}_{xyyx}(E^T_0)^2E^\omega_x\\
                                                    \chi^{(3)}_{xyyx}(E^T_0)^2E^\omega_y\\
                                                    0
                              \end{matrix}
                      \right).
\end{eqnarray}
Under these conditions, the new principal axes of the refractive index ellipsoid correspond to the directions [$110$], [$\bar110$], and [$001$], which we label as $x'$, $y'$, and $z'$.  The third order polarization in these new coordinates becomes
\begin{eqnarray}
\nonumber
\vec{P}'^{(3)}=\left( \begin{matrix} 0\\ 
                                                     \chi^{(3)}_{xyyx}(E^T_0)^2E^\omega_{y'}\\
                                                     0
                              \end{matrix}
                      \right).
\end{eqnarray}
The gating beam $E^\omega$ is polarized along the principal $y'$ axis and does not experience polarization rotation due to the THz Kerr effect. Thus, $\Delta I(t)$=0 for  $\alpha$=$0\dg$.  In the case $\alpha$=$90\dg$, we set
\begin{eqnarray}
\nonumber
\vec{E}^T=\frac{E^T_0}{\sqrt{2}}\left( \begin{matrix} 1\\ -1\\ 0\\ \end{matrix} \right) \; \textrm{and} \quad \vec{E}^{\omega}=E^{\omega}_0\left( \begin{matrix}0\\ 0\\ 1\\ \end{matrix}\right)\\ 
\nonumber
\textrm{and obtain} \quad
\vec{P}^{(3)}=\left(
                              \begin{matrix}0\\
                                                    0\\
                                                    \chi^{(3)}_{xyyx}(E^T_0)^2E^{\omega}_z
                              \end{matrix}
                      \right).
\end{eqnarray}
The [$001$] $z$ direction remains a principal axis of the refractive index ellipsoid, and the gating beam remains polarized along $z$.  Thus, $\Delta I(t)$=0 for  $\alpha$=$90\dg$. 

Our measurements confirmed that $\Delta I(t)$=0 for $\alpha$=$0\dg$ and $90\dg$ when detected at the second harmonic, correspondig to the THz Kerr effect.  When detected at the first harmonic, $\Delta I(t)$ is zero for $\alpha$=$0\dg$, but for $\alpha$=$90\dg$ it exhibits a maximum, which is the position of the most efficient linear electro-optic detection of the THz waveform\cite{planken:313}.  The data presented in Figs.~\ref{fig:tdsgap},\ref{fig:fdsgap} were collected near the angle $\alpha=70\dg$, close to the maximum of first harmonic balanced photodiode signal $\Delta I(t)$.

We now show that the second harmonic lock-in detection can be used to measure THz second harmonic generation in noncentrosymmetric media due to the second order polarizability $P^{(2)}_i=\chi^{(2)}_{ijk}E^T_jE^T_k$.  In this measurement, the THz pulses are focused into a sample generating the THz second harmonic after passing through the fixed wire grid polarizer, Fig.~\ref{fig:fig1}.  The generated second harmonic THz pulse $E^{T(2)}$ propagates together with the fundamental THz pulse $E^T$ after passing through the sample.  This propagating THz wave $E^{T(2)}$ can be detected via the linear electro-optic effect in the GaP receiver.  In a conventional THz time-domain spectrometer without the polarizing chopper and sinusoidal modulation, it is difficult to separate the contributions of $E^{T(2)}$ and $E^{T}$ to the linear electro-optic effect, since the two pulses co-propagate with significant overlap in both time and frequency domains.  In addition, the expected amplitude of the generated second-harmonic pulse $E^{T(2)}$ may be several orders of magnitude smaller than the fundamental amplitude $E^T$\cite{merbold:7262}.  The sinusoidal modulation by the polarizing chopper allows the separation of the two waves, as the second-harmonic wave $E^{T(2)}$ is modulated at twice the sinusoidal modulation frequency, Fig.~\ref{fig:fig1}.  Then, the fundamental THz wave $E^T$ is detected at the fundamental modulation frequency and the second-harmonic wave $E^{T(2)}$ is measured by the lock-in amplifier at the second harmonic of the sinusoidal modulation frequency.  We emphasize that both waves are measured via the linear electro-optic Pockels effect in GaP, but are separated via the first and second harmonic lock-in detection.  To distinguish the linear electro-optic effect due to $E^{T(2)}$ and the quadratic THz Kerr effect due to $E^T$ in the GaP receiver, which are both measured in the second harmonic lock-in detection, we must eliminate the latter by choosing the angle $\alpha=90\dg$ in the configuration of Fig.~\ref{fig:fig4}.  In this geometry, the THz Kerr effect is zero and the electro-optic Pockels effect is maximized\cite{planken:313}.  This scheme should allow the measurement of THz second harmonic generation that so far has eluded researchers.  Theoretical studies of possible materials for observing the phenomenon have been rare.  Merbold, Bitzer, and Feurer have considered the use of metal nanoslits and split-ring resonators to enhance the peak THz field and the nonlinear interactions\cite{merbold:7262}.  They predict that THz second harmonic generation could potentially be observed in LiTaO$_3$ with the help of THz field enhancement by the nanoslits or the split-ring resonators.  

\section{Conclusions}
In summary, we have presented a measurement of the quadratic THz Kerr effect in the (110) oriented zincblende crystal GaP.  In our measurement geometry, the THz Kerr birefringence is superimposed on the much stronger birefringence due to the linear electro-optic effect.  We used the sinusoidal THz pulse train modulation and second harmonic lock-in detection to separate and measure the much weaker THz Kerr birefringence.  More generally, our measurement method allows a separation of the electro-optic signals proportional to $E^T$ and $(E^T)^2$, as evidenced by the presented data.  We predict that our method will allow the measurement of the elusive THz second harmonic generation in noncentrosymmetric media, which opens a new area of exploration in nonlinear optics and optical properties of materials - the second harmonic THz spectroscopy and related photonic technologies.  

\section*{Acknowledgments}
The work at Tulane University was supported by the NSF Award No. DMR-1554866 and by the Carol Lavin Bernick Faculty Grant Program.  


\end{document}